\documentclass{article}

\usepackage{microtype}
\usepackage{graphicx}
\usepackage{subfigure}
\usepackage{booktabs} 

\usepackage{hyperref}

\newcommand{\phiDOA}{\phi_{0}}
\newcommand{\thetaDOA}{\theta_{0}}
\newcommand{\com}[1]{``#1''}
\usepackage{optidef}
\usepackage{scalerel}
\graphicspath{{images}}
\usepackage{siunitx}

\usepackage[accepted]{icml2025}

\usepackage{amsmath}
\usepackage{amssymb}
\usepackage{mathtools}
\usepackage{amsthm}

\usepackage{booktabs} 
\usepackage{multirow} 
\usepackage{array}    

\usepackage{amsfonts}

\usepackage[capitalize,noabbrev]{cleveref}

\theoremstyle{plain}

\theoremstyle{definition}

\theoremstyle{remark}

\usepackage[textsize=tiny]{todonotes}

\icmltitlerunning{Frequency-Invariant Beamforming in Elevation and Azimuth via Autograd and Concentric Circular Microphone Arrays}

\begin{document}

\twocolumn[
\icmltitle{Frequency-Invariant Beamforming in Elevation and Azimuth via Autograd and Concentric Circular Microphone Arrays}



\icmlsetsymbol{equal}{*}

\begin{icmlauthorlist}
\icmlauthor{Jorge Ortigoso-Narro}{aaa}
\icmlauthor{Jose A. Belloch}{aaa}
\icmlauthor{Maximo Morales-Cespedes}{bbb}
\icmlauthor{Maximo Cobos}{ccc}
\end{icmlauthorlist}

\icmlaffiliation{aaa}{Department of Electronic Technology, Universidad Carlos III de Madrid, Spain}
\icmlaffiliation{bbb}{Department of Signal Theory and Communications, Universidad Carlos III de Madrid, Spain}
\icmlaffiliation{ccc}{Department of Computer Science, Universidad de Valencia, Spain}

\icmlcorrespondingauthor{Jorge Ortigoso-Narro}{jortigos@pa.uc3m.es}

\icmlkeywords{Beamforming, Autograd, Concentric Circular Arrays, Frequency Invariant.}

\vskip 0.3in
]



\printAffiliationsAndNotice{Preprint, sumbitted to Acta Acustica from Forum Acusticum 2025.}  

\begin{abstract}
The use of planar and concentric circular microphone arrays in beamforming has gained attention due to their ability to optimize both azimuth and elevation angles, making them ideal for spatial audio tasks like sound source localization and noise suppression. Unlike linear arrays, which restrict steering to a single axis, 2D arrays offer dual-axis optimization, although elevation control remains challenging. This study explores the integration of autograd, an automatic differentiation tool, with concentric circular arrays to impose beamwidth and frequency invariance constraints. This enables continuous optimization over both angles while maintaining performance across a wide frequency range. We evaluate our method through simulations of beamwidth, white noise gain, and directivity across multiple frequencies. A comparative analysis is presented against standard and advanced beamformers, including delay-and-sum, modified delay-and-sum, a Jacobi-Anger expansion-based method, and a Gaussian window-based gradient descent approach. Our method achieves superior spatial selectivity and narrower mainlobes, particularly in the elevation axis at lower frequencies. These results underscore the effectiveness of our approach in enhancing beamforming performance for acoustic sensing and spatial audio applications requiring precise dual-axis control.
\end{abstract}

\section{Introduction}
\label{sec:introduction}
Acoustic beamforming is a spatial filtering technique that processes signals captured by a microphone array to isolate sounds arriving from a specific direction. This is achieved by introducing precise phase shifts or time delays to the signals recorded by each microphone, effectively steering the array's sensitivity toward the desired direction while attenuating noise and interference from other directions. By exploiting the geometric arrangement of the array and the direction of arrival (DoA) of the sound waves, beamforming enhances the signal-to-noise ratio (SNR) and facilitates spatial discrimination.

In the context of acoustic applications, beamforming plays a critical role in enhancing speech intelligibility, localizing sound sources, and suppressing ambient noise \cite{source_localization}. More complex applications include acoustic cameras \cite{64MICMODULEOWN}, ultrasound sensing networks or beamfoming-sonar applications. However, challenges such as mitigating the effects of reverberation, handling low SNR conditions, and balancing computational complexity with real-time performance remain active areas of research.

Planar microphone arrays enable 3D beamforming, offering independent control over azimuth and elevation, which is essential for applications like aerial sensing and immersive audio. Unlike linear arrays limited to horizontal steering, planar arrays achieve vertical steering using phase shifts to compensate for microphone height differences. However, effective elevation steering faces challenges. Limited vertical aperture in compact arrays restricts resolution, especially for low frequencies. Moreover, it's highly sensitive to vertical microphone placement errors, which can cause significant phase issues. Non-uniform vertical microphone distribution can also lead to asymmetrical elevation beam-patterns with uneven sidelobes and reduced spatial selectivity.

Numerous studies have explored methods to achieve frequency invariance in microphone arrays. A common approach employs differential arrays, which leverage spatial derivatives of the sound field to create directional beampatterns \cite{diff_theory1, diff_theory_book}. However, the requirement for small inter-sensor spacing in these arrays limits their bandwidth and compromises low-frequency performance. To overcome these challenges, concentric circular microphone arrays (CCMAs) have emerged as a compelling alternative. Their symmetric design enables not only frequency-invariant beamforming across a wide frequency range but also improves steering capabilities. Recent investigations have even combined differential beamforming techniques with CCMAs to further exploit these advantages \cite{ccma_steer1, ccma_diff, Parra2006}.

Other strategies to attain frequency invariance include iterative optimization techniques that derive weighting coefficients without relying on strict differential constraints \cite{israkroneken, isragaussian, isra3}. Although automatic differentiation appears suitable, its adoption in steerable acoustic beamforming remains limited. In \cite{autograd_beamformer_design}, the authors demonstrated how automatic gradient differentiation could be integrated to optimize linear differential microphone arrays (LDMAs).

In this work, we address these limitations by proposing a novel beamforming framework for concentric circular microphone arrays based on automatic differentiation. The proposed approach enables the simultaneous control of both azimuth and elevation beamwidths, while ensuring frequency-invariant performance and robustness to spatially uncorrelated noise. Unlike previous methods, our framework incorporates a differentiable objective function that combines directivity metrics with additional regularization terms, facilitating a flexible and effective optimization process. The effectiveness of the proposed method is validated through comprehensive simulations, demonstrating its advantages over conventional and state-of-the-art techniques.

\section{Signal model and problem formulation} 
In this work, we consider concentric circular microphone arrays composed of $R$ rings, each uniquely indexed by $r = \{1, 2, \dots, R\}$. Each ring is defined by its radius $\rho_r$, which determines the spatial distribution of the microphones along that ring. Furthermore, each ring contains $M_r$ microphones; each microphone is characterized by the common radial distance $\rho_r$ and its unique angular position $\phi_{r,m}$, where $m = \{1, \dots, M_r\}$ denotes the microphone’s index and $\phi_{r,m}$ is measured relative to the positive x-axis. In addition, the Direction of Arrival (DoA) of a sound source is specified by two angles: the \emph{azimuth angle} $\phi_{0}$, measured in the plane parallel to the array (from the x-axis), and the \emph{elevation angle} $\theta_{0}$, measured from the horizontal plane.
\begin{figure}[b]
    \includegraphics[width=7.8cm, trim=0cm 0cm 0cm 0.1cm, clip]{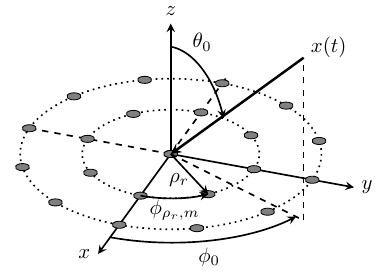}
    \vspace{-3mm}
    \caption{Problem geometry.}
    \label{fig:signal_model}
    \vspace{-3mm}
\end{figure}

Given a concentric circular microphone array as shown in Fig.~\ref{fig:signal_model}, let $x(t)$ be an incident wave arriving from the direction $(\phi_{0}, \theta_{0})$. As shown in Eq.~\ref{eq:rx}, each microphone records a delayed version of $x(t)$,
\begin{equation} \label{eq:rx}
    y_{r, m}(t) = x(t - \tau_{\mu_{r, m}}) + n_{r,m},
\end{equation}
where $n_{r,m}$ represents the noise and distortion acquired by each microphone. The delay, which depends on the microphone's position relative to the central microphone is defined in Eq.~\ref{eq:tau}:
\begin{equation} \label{eq:tau}
    \tau_{\mu_{r,m}} = -f_s\frac{\rho_r}{c}\sin\left(\theta_{0}\right) \cos\left(\phi_{0} - \phi_{r, m}\right),
\end{equation}
with $f_s$ as the sampling frequency and $c$ as the speed of sound. In the frequency domain, the model becomes
\begin{align} \label{eq:freq_domain}
\begin{split} 
    Y_{r,m}(f) &= X(f) \, e^{-j 2\pi f (-\tau_{r, m})} + N(f)\\
    &= X(f) \, d_{r, m}(f, \theta_{0},\phi_{0}) + N(f),
\end{split}
\end{align}
where $d(f, \theta, \phi)$ denotes the steering vector at orientation $(\theta, \phi)$. By combining these expressions, the overall processed signal can be written in matrix form (see Eq.~\ref{eq:matrix_form}). Here, $\mathbf{h}(f), \mathbf{d}(f), \mathbf{y}(f), \mathbf{n}(f) \in \mathbb{C}^{M_T \times 1}$, where $M_T$ is the total number of microphones and $B$ is the number of frequency bands.
\begin{equation} \label{eq:matrix_form}
    \mathbf{y}(f) = x(f) \, \mathbf{d}(f, \theta, \phi) + \mathbf{n}(f)
\end{equation}
\begin{equation} \label{eq:z_mf}
    \mathbf{z}(f) = \mathbf{h}(f)^H \left( x(f) \, \mathbf{d}(f, \theta, \phi) + \mathbf{n}(f) \right).
\end{equation}
The beamformer coefficients are grouped inside $\mathbf{h}$. To be able to recover the original signal without distortion, the desired filter response can be obtained by finding $\mathbf{h}(f)$ so that $\mathbf{h}(f)^H\mathbf{d}(f, \theta, \phi) = 1$.

\subsection{Evaluation metrics}

To assess the performance of the microphone array, directivity pattern-related metrics are analyzed. First, the 3D-beampattern, described in Eq.~\ref{eq:beampattern} represents the gain of the array as a function of both azimuth and elevation.
This pattern not only illustrates the primary lobe, where the array's response is maximized, but also highlights the sidelobes, which can be potential sources of interference from undesired directions. The $\SI{-3}{\dB}$ elevation and azimuth beamwidths can be obtained by solving the following equations:
\begin{equation} \label{eq:beampattern}
    B(f, \theta, \phi) = \mathbf{h}(f)^H \mathbf{d}(f, \theta, \phi),
\end{equation}
\begin{equation} \label{eq:elevation_beamwidth}
    \Theta: \frac{|B(f, \theta, \phiDOA)|}{|B(f, \thetaDOA, \phiDOA|} = \frac{1}{2},
\end{equation}
\begin{equation} \label{eq:azimuth_beamwidth}
    \Phi: \frac{|B(f, \thetaDOA, \phi)|}{|B(f, \thetaDOA, \phiDOA|} = \frac{1}{2}.
\end{equation}
Complementing this spatial analysis, the directivity index, described in Eq.~\ref{eq:directivity} offers a quantitative measure by comparing the gain in the desired direction to the average gain across all directions, effectively summarizing the array’s ability to focus on the target signal. 
\begin{align} \label{eq:directivity}
    &\mathrm{DF}(f) = \\ &\frac{|B(f, \thetaDOA, \phiDOA)|^2}{\frac{1}{4\pi} \int_0^\pi \int_0^{2\pi} |B(f, \theta, \phi)|^2 d\phi d\theta} = \frac{|\mathbf{h}(f)^H \mathbf{d}(f, \thetaDOA, \phiDOA)|^2}{\mathbf{h}(f)^H \Gamma(f) \mathbf{h}(f)} \nonumber \\
    &\Gamma_{i,j}(f) = \text{sinc}\left(\frac{2\pi f l_{i,j}}{c}\right) \nonumber
\end{align}
where $l_{i,j}$ represents the Euclidean distance between the microphones positioned at $i$ and $j$ respectively. The white noise gain metric (WNG), can be defined as
\begin{equation} 
\label{eq:wng}
    \mathrm{WNG}(f) = \frac{|\mathbf{h}(f)^H \mathbf{d}(f, \theta_0, \phi_0)|^2}{\mathbf{h}(f)^H \mathbf{h}(f)},
 \end{equation}
which can be employed for measuring the SNR enhancement in presence of spatially uncorrelated noise, serving as a critical indicator of the beamformer’s noise suppression capabilities. Together, all these metrics provide a robust framework for evaluating and optimizing the acoustic beamforming performance of the microphone array.

\subsection{Beamformer design}
To place the microphones across the rings, a non-aliasing constraint is assumed to select the minimum distance from microphones. By selecting a minimum chord of length $\lambda_\mathrm{min}/2$, the delay between elements covers at least half-wavelength of the maximum frequency. As a consequence, the minimum number of microphones in each ring and their respective angular positions are given by
\begin{equation} \label{eq:Mr}
    M_r = \left\lfloor\frac{\pi}{\arcsin\left(\lambda_\mathrm{min}/(4\rho_r)\right)}\right\rfloor,
\end{equation}
and
\begin{equation} \label{eq:mics_phi}
    \phi_{r,m} = 2\pi \frac{m}{M_r}, \quad  m = \{0, 1, \dots, M_r-1\},
\end{equation}
respectively. To control the contribution of each ring to the overall beamformer, a set of frequency-dependent ring-level weights is introduced, where each weight satisfies $0 \leq w_r(f) \leq 1$. These weights determine the relative importance of each ring in shaping the beampattern and are normalized such that $\sum_{r=1}^{R} w_r(f) = 1$. Then, following \cite{isragaussian, windowbased}, a Gaussian window is used to weight the microphones of each ring, giving more weight to those more aligned with the DoA. The intra-ring weights can be computed as
\begin{equation} \label{eq:gaussian_window}
    s_{r,m} = \mathrm{GW}[{0, \sigma_{r}^{2}]}(\delta_{r,m}), 
\end{equation}
where $\delta_{r, m}$ can be obtained by calculating the vector distance given by
\begin{align} \label{eq:vect_dist}
    \delta_{r,m} &= \left\lVert\mathbf{v}_{\pi/2, \phi_{r,m}} - \mathbf{v}_{\thetaDOA, \phiDOA}\right\lVert_2 \\
    \delta_{r,m} &:= \left(\delta_{r,m} - \min(\delta)\right) / \max(\delta) \label{eq:delta_norm}
\end{align}
To ensure robust angular distance calculations, the angles must be wrapped so that when $\phi_{r,m} \notin [\phiDOA - \frac{\pi}{2}, \phiDOA + \frac{\pi}{2}]$ the calculations simply add $\pi$ to the DoA angles. Then, $\delta$ gets normalized as described in Eq.~\ref{eq:delta_norm}. By combining the two weighting coefficients $w$ and $s$, the final filter coefficient $h$ can be calculated as
\begin{equation} \label{eq:h}
    h_{r,m} = w_r(f) s_{r,m}(f) d_{r,m}(f, \thetaDOA, \phiDOA).
\end{equation}

\section{Optimization}
To determine the optimal beamforming coefficients, an objective loss function is minimized using Autograd, an automatic differentiation framework \cite{paszke2017automatic}. Internally, Autograd constructs a dynamic computational graph that records all tensor operations, enabling efficient application of the chain rule via reverse-mode differentiation. This approach eliminates manual gradient computation, ensuring both accuracy and efficiency in optimizing the beamforming parameters. Moreover, the framework's modular design simplifies the testing of various loss functions by allowing easy modification or combination of different loss components. This flexibility facilitates rapid experimentation and evaluation of alternative formulations to better address specific system requirements.

The loss function can be constructed by combining different terms to enhance various characteristics among the system's performance metrics. For example, one may combine a term that penalizes signal distortion with another that limits interference leakage, while also incorporating power constraints to manage energy efficiency. Regularization terms can be added to promote smoothness, ensuring robust performance across different operating conditions. By assigning appropriate weights to these terms, the composite loss function can be finely tuned to emphasize desired attributes such as beam sharpness, reduced sidelobe levels or improved overall system stability, thereby enabling a balanced optimization of the beamforming coefficients tailored to specific requirements. Formally, this approach corresponds to solving the following optimization problem,
\begin{mini}
    {w, \sigma}{\mathcal{L}_f(w, \sigma)}{}{\textbf{P1}: \quad }
    \addConstraint{\Theta_f(w, \sigma)}{\leq \Theta_{BW}}
    \addConstraint{\Phi_f(w, \sigma)}{\leq \Phi_{BW}}
    \addConstraint{\mathbf{1}^T w}{= 1}
    \addConstraint{0 \preceq w}{\preceq 1}
    \addConstraint{\sigma}{\succeq 0}
    \label{eq:optimization_problem}
\end{mini}
where $\mathcal{L}_f$ is the loss function evaluated at the frequency $f$, $\Theta_f$ and $\Phi_f$ are the $\SI{-6}{\dB}$ beamwidths in elevation and azimuth, respectively.

\subsection{Differentiabily and implementation details}

To optimize functions that are not inherently differentiable using gradient descent, specific modifications and techniques must be applied. For example, \eqref{eq:elevation_beamwidth} and \eqref{eq:azimuth_beamwidth} require the $\arg \min$ operation over a sampled grid and tensor indexing, both of which are non-differentiable operations. Instead, differentiable approximations such as temperature-modified softmax functions or sigmoid masks can be applied to mimic this behavior in a soft-differentiable manner.

\begin{figure}[ht]
 \includegraphics[width=7.8cm, keepaspectratio]{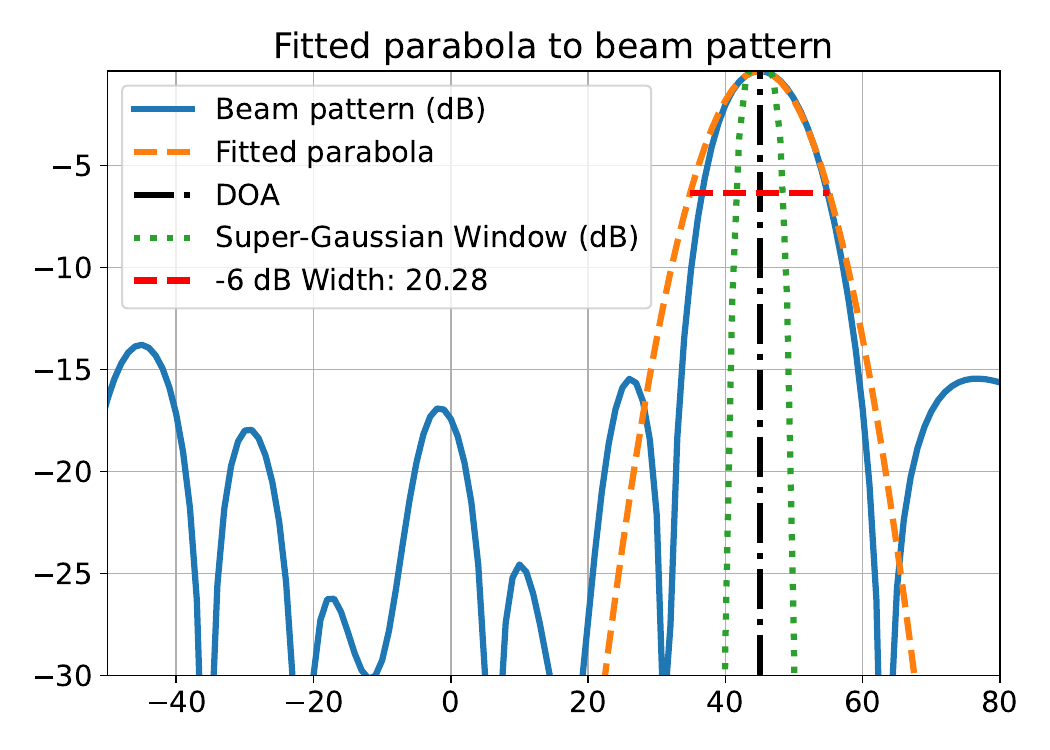}
 \vspace{-3mm}
 \caption{Fitted parabola to the beampattern.}
  \vspace{-3mm}
 \label{fig:parabola_fit}
\end{figure}

To determine the $\SI{-6}{\dB}$ points over the grid, softmax approaches were found to be inadequate due to insufficient gradient propagation at steep edges. Consequently, the final beamwidth calculation method involved applying a weighted least-squares fit to the beampattern. This approach fits a parabolic model directly to the beampattern parameters, thereby enabling a direct estimation of the beamwidth. To prevent the quadratic fit from being unduly influenced by sidelobes, a super-Gaussian mask was applied around the DoA to weight the samples and mitigate the effects of poor sidelobe fitting. Furthermore, to compensate for variations in beamwidth across different frequency bands, the window width was adjusted to be narrower at higher frequencies by selecting a span smaller than the mainlobe to improve the parabola fit. 
The procedure for estimating the beamwidth in elevation, with an equivalent procedure applied to the azimuthal axis, can be written as
\begin{equation}
\Theta = 2 \sqrt{\frac{\Delta L}{|a|}}
\end{equation}
where $a = \frac{W \cdot S_B - S_x \cdot B}{W \cdot S_x^2 - S_x^2}$,  $W = \sum_i w_{i,\theta}$ and $S_x = \sum_i w_{i,\theta} x_i^2$, and $S_B = \sum_i w_{i,\theta} x_i^2 B_i$ with $B = \sum_i w_{i,\theta} B_i$, $w_{i,\theta} = \exp(-\frac{1}{2} \left(\theta_i - \thetaDOA/\sigma_\theta\right)^4)$ and $\quad x_i = \theta_i - \thetaDOA$.  Moreover, $\theta_i$ are the discrete elevation angles, $\sigma_\theta$ is the standard deviation of the window (which may vary with frequency), $B_i$ denotes the beampattern value in dB at $\theta_i$, and $\Delta L$ is the power-level difference, e.g., $\SI{-6}{\dB}$. An example of the resulting fit and beamwidth estimation is shown in Fig.~\ref{fig:parabola_fit}.

The framework employed to implement the experiments was PyTorch 2.6.0 \cite{paszke2017automatic}. A learning rate of $\eta = 0.1$ was used in combination with the RProp \cite{RProp} gradient-descent algorithm with parameters $(\eta^+, \eta^-) = (0.5, 1.2)$, $(\Gamma_{\max}, \Gamma_{\min}) = (10^{-6}, 50)$.

\section{Experiments and results}

\begin{figure*}[!htpb]
    \includegraphics[width=\textwidth]{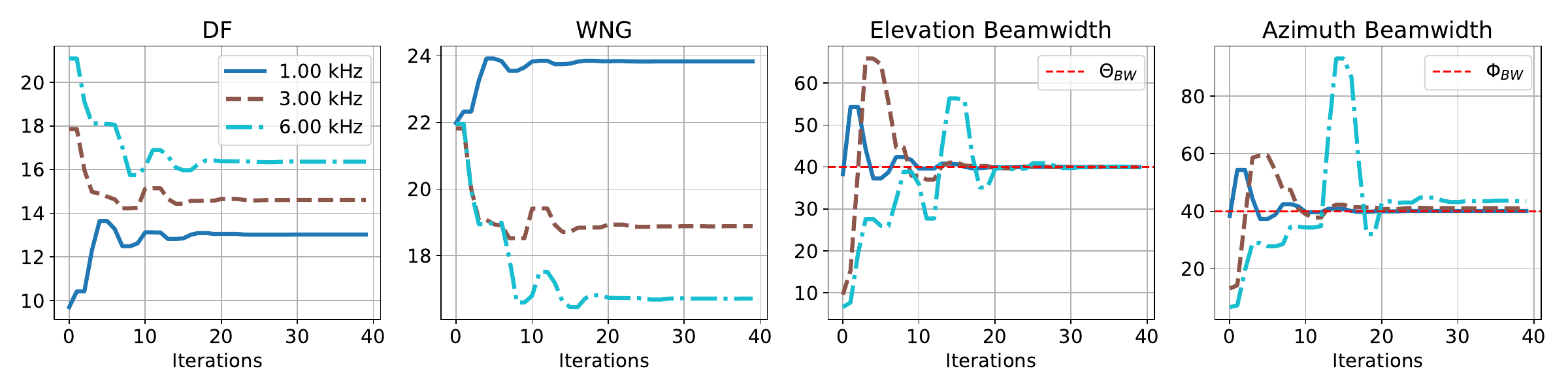}
    \vspace{-6mm}
    \caption{Performance evolution with the iterations of the base array optimized with Eq.~\ref{eq:loss_1} ($\mathcal{L}_1$) as objective.}
    \label{fig:metrics_vs_iters}
\end{figure*}

In the main experiments, we consider an array of 5 rings with radii $r = \{0, 5, 10, 15, 20\}$ cm and a sampling frequency of $f_s = \SI{16}{\kHz}$. Figure~\ref{fig:metrics_vs_iters} shows how the optimization process takes place by using $\mathcal{L}_1$ as the loss function. The objective beamwidths selected were $\Theta_{BW}=\Phi_{BW}=\SI{40}{\degree}$ and the DOA was chosen to be $(\SI{45}{\degree}, \SI{45}{\degree})$.
\begin{figure}[t!]
    \includegraphics[width=8.5cm]{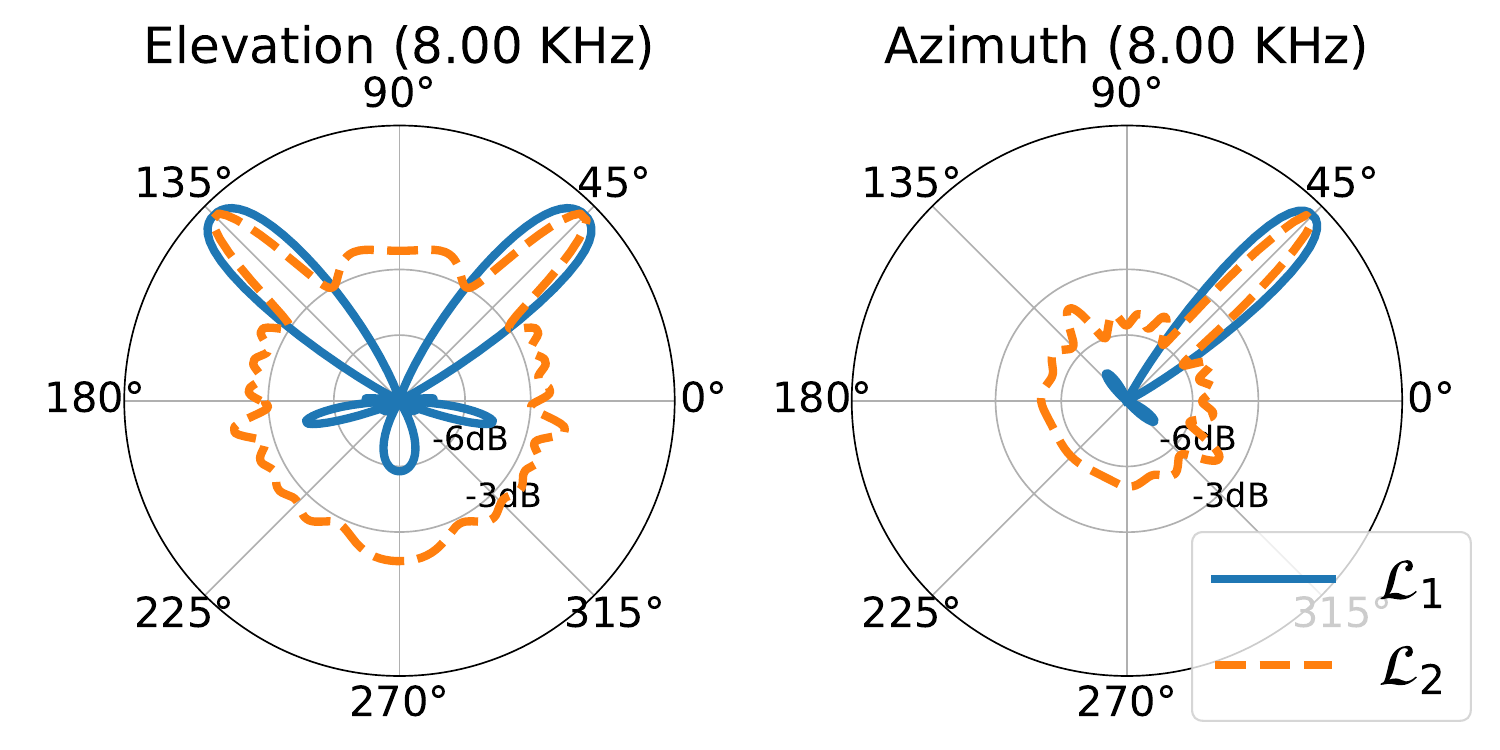}
    \caption{Beampattern comparison by correcting beamwidths smaller than the objective by explicitly increasing the estimated mainlobe beamwidth ($\mathcal{L}_1$) vs reducing the directivity factor ($\mathcal{L}_2$).}
    \label{fig:b_vs_minD}
\end{figure}

\subsection{Objective functions}
Leveraging the flexibility of the implemented framework, several objective functions were tested with the goal of promoting frequency invariance. First, the approach suggested in \cite{isragaussian, israkroneken} was adopted, where an objective beamwidth is employed in combination with the array's directivity as a loss metric (see Eq.~\ref{eq:loss_1}). 
\begin{equation} \label{eq:loss_1}
    \mathcal{L}_{1} = 
    \begin{cases}
        \Theta_f, & \Theta_f > \Theta_\mathrm{BW} \ \text{\&} \ \Phi_f \leq  \Phi_\mathrm{BW} \\
        \Phi_f, & \Theta_f \leq \Theta_\mathrm{BW} \ \text{\&} \ \Phi_f >  \Phi_\mathrm{BW} \\
        - \log_{10} \mathrm{DF}, & \text{otherwise} \\
    \end{cases}
\end{equation}

In an attempt of improving the array's performance, the white noise gain (WNG) was also included with a trade-off coefficient to complementarily weight both $\mathrm{DF}$ and $\mathrm{WNG}$ as suggested in \cite{isra3, tradeoff_kroneken_constant}. The $\mathcal{L}_1$ function maximizes the array’s directivity when the beamwidth exceeds the desired target, while it simultaneously promotes a broader beam when the beamwidth is too narrow. This strategy maintains a constant beamwidth without requiring an explicit reduction in the directivity factor for frequencies that yield a narrow main lobe, thereby preventing the beamformer from developing large sidelobes. Figure~\ref{fig:b_vs_minD} illustrates this behavior by comparing the high-frequency beampattern obtained using $\mathcal{L}_1$ with that produced by a new $\mathcal{L}_2$ approach, which instead reduces the directivity factor to correct for beamwidth discrepancies.

In this study, we also explore the inclusion of additional terms to explicitly promote performance invariance across the considered frequency range, although this might affect negatively to the best-working bands. In particular we explored the inclusion of terms to penalize the standard deviation of the metrics across the frequency range and the differences between adjacent and opposing frequency bands, the complete tunable loss function is shown in Eq.~\ref{eq:loss_3}.
\begin{align} \label{eq:loss_3}
    \mathcal{L}_{3} &= 
    \begin{cases}
        \Theta_f, & \Theta_f > \Theta_{BW} \ \text{\&} \ \Phi_f \leq  \Phi_{BW} \\
        \Phi_f, & \Theta_f \leq \Theta_{BW} \ \text{\&} \ \Phi_f >  \Phi_{BW} \\
        P + I + \Delta & \text{otherwise}
    \end{cases}, \\ 
     P &= - \alpha\log_{10} \mathrm{DF} - (1-\alpha)\log_{10} \mathrm{WNG}, \label{eq:P}\\
     I &= \lambda_1 \ \mathrm{std}(\mathrm{DF}) + \lambda_2 \ \mathrm{std}(\mathrm{WNG}), \label{eq:I}\\
     \Delta &= \lambda_3\textstyle \sum_{i=2}^{\left \lfloor{F/2}\right \rfloor } |P_{i} - P_{N-i+1}| \label{eq:delta}
\end{align}
where $P$ (Eq.~\ref{eq:P}) is the performance term, $I$ (Eq.~\ref{eq:I}) the invariance term, and $\Delta$ (Eq.~\ref{eq:delta}) the differences term. $\lambda_{1}$, $\lambda_{2}$, and $\lambda_{3}$ control the weight each term is given while $\alpha$ is the trade-off coefficient between directivity factor and white noise gain. Note that $\mathcal{L}_1$ can be achieved by choosing $\alpha=1, \lambda_1 = 0, \lambda_2 = 0$ and $\lambda_3 = 0$.

\begin{figure*}[!htpb]
    \centering
    \includegraphics[width=\textwidth]{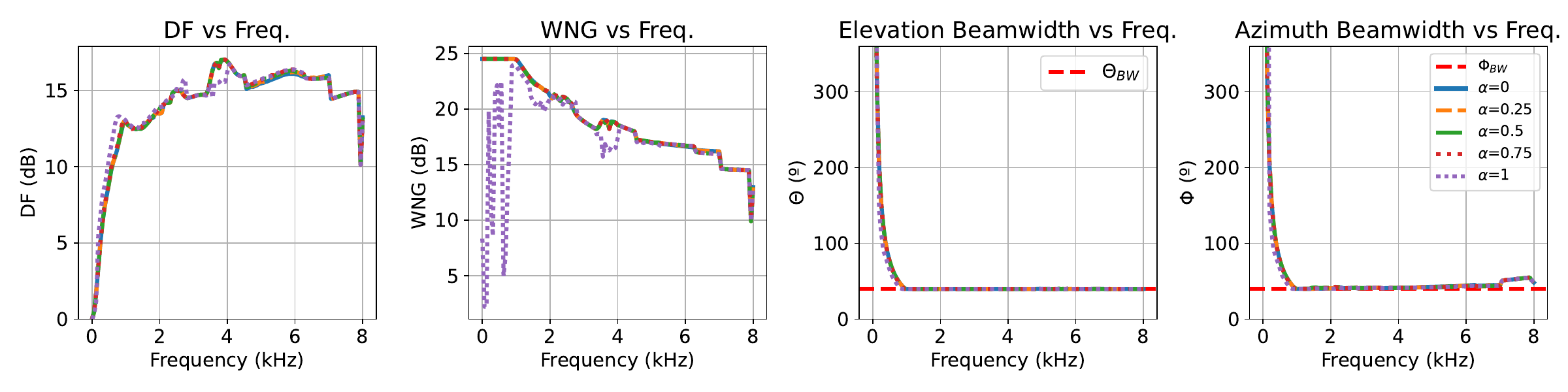} \\
    \includegraphics[width=\textwidth]{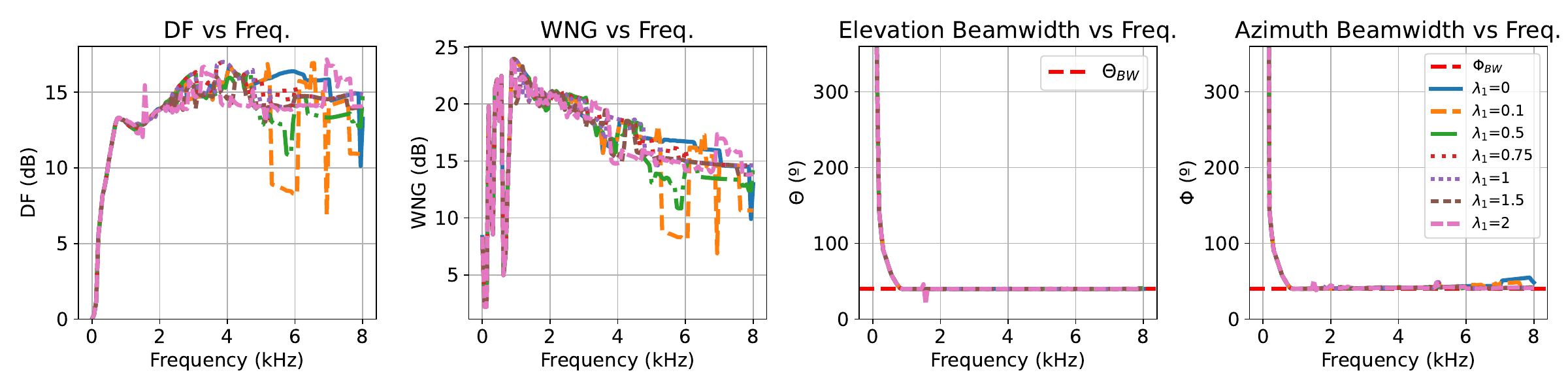} \\
    \includegraphics[width=\textwidth]{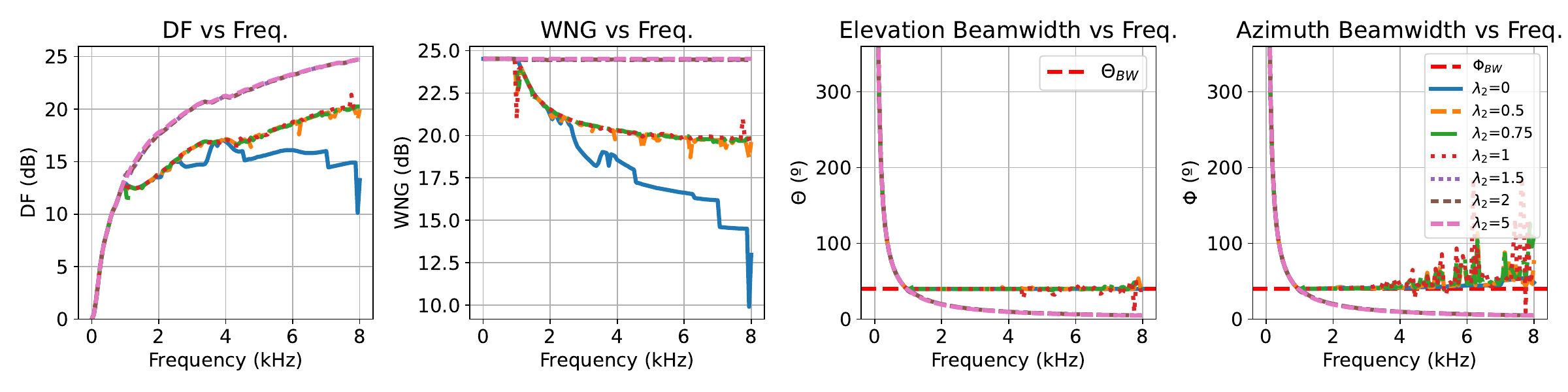} \\
    \includegraphics[width=\textwidth]{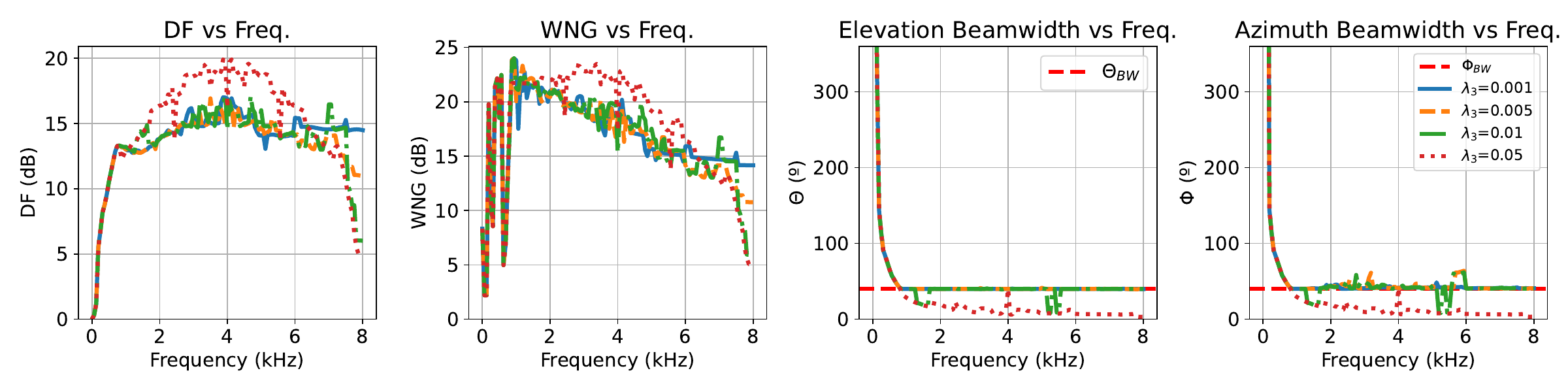} \\
    \vspace{-3mm}
    \caption{Results obtained for different parametrization of $\mathcal{L}_3$ in terms of metrics s frequency. Top row shows $\alpha=\{0, 0.25, 0.5, 0.75, 1\}$, $\lambda_1 = \lambda_2, \lambda_3 = 0$. Second row shows the effect of varying $\lambda_1 = \{0, 0.1, 0.5, 0.75, 1, 1.5, 2\}$ with $\alpha=1, \lambda_2=1$ and $\lambda_3=0$. The third row sweeps $\lambda_2 = \{0, 0.5, 0.75, 1, 1.5, 2, 5\}$ for $\alpha = 0, \lambda_1 = 0$ and $\lambda_3 = 0$. Finally the last row evaluates $\lambda_3 = \{0.001, 0.005, 0.01, 0.05\}$ while fixing $\alpha=1$, $\lambda_1$ and $\lambda_2=0$.}
    \label{fig:l3}
\end{figure*}

Figure~\ref{fig:l3} presents the results for various $\mathcal{L}_3$ configurations. In the top row, the metrics are compared for different $\alpha$ values, which serve as a tradeoff between the directivity factor and the white noise gain. The results indicate that $\alpha$ has minimal impact on the final outcome, as its role is limited to adjusting higher beamwidth corrections due to the piecewise nature of the function.

In the second and third rows, several values of $\lambda_1$ and $\lambda_2$ are examined to demonstrate the effect of incorporating a penalization term on the spread (standard deviation) of the metrics across the frequency range of interest. Although this term did not significantly affect the results in the first experiment, subsequent experiments show that optimizing with the white noise gain and including the regularization term can achieve a flatter white noise gain without substantially compromising directivity invariance. Finally, the last row illustrates the results for different values of $\lambda_3$ for the considered best-performing settings ($\alpha=1, \lambda_1 = 1, \lambda_2=0$), it can be seen how for small values a flatter directivity factor curve is obtained.

\subsection{Comparison with other methods}
\begin{figure}[!htpb]
    \centering
     \includegraphics[width=8cm]{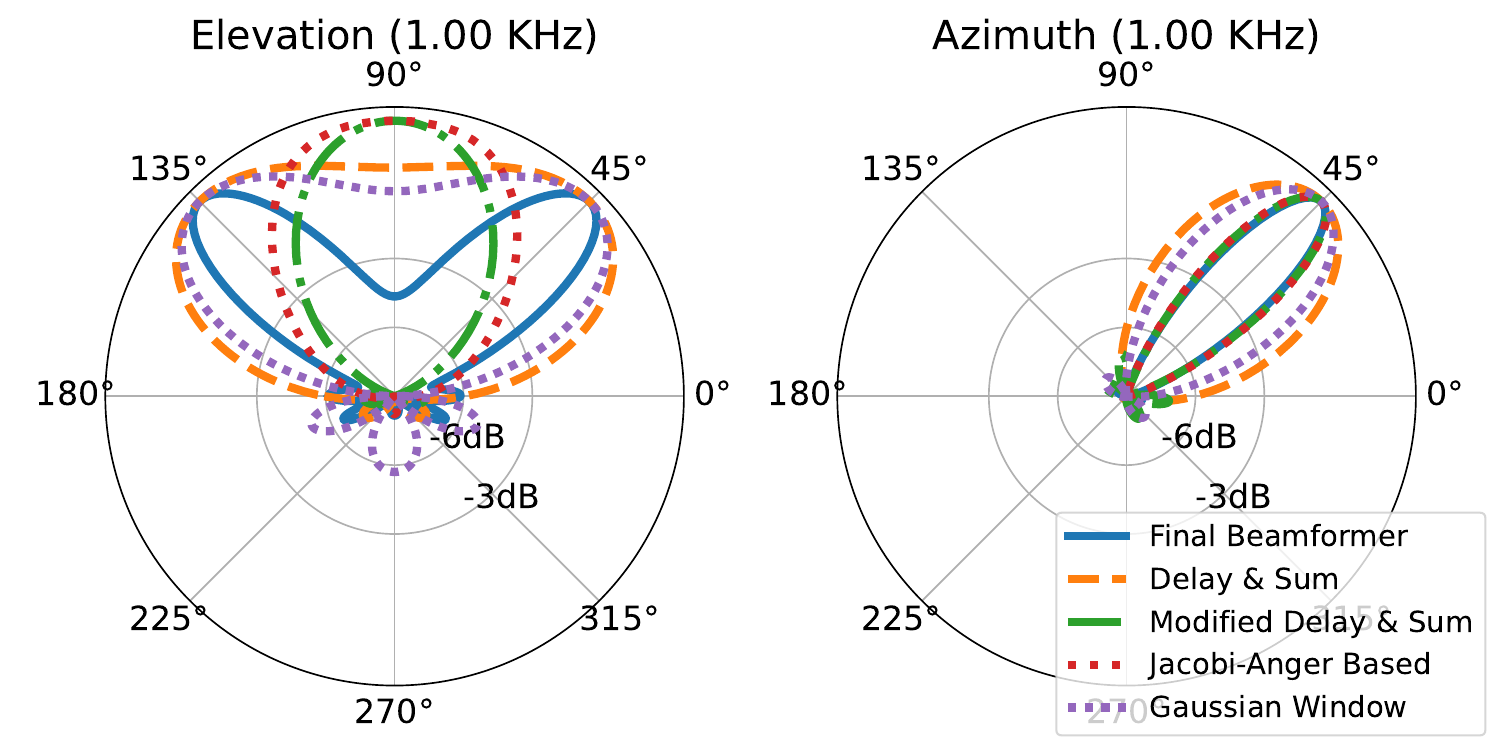} \\
     \includegraphics[width=8cm]{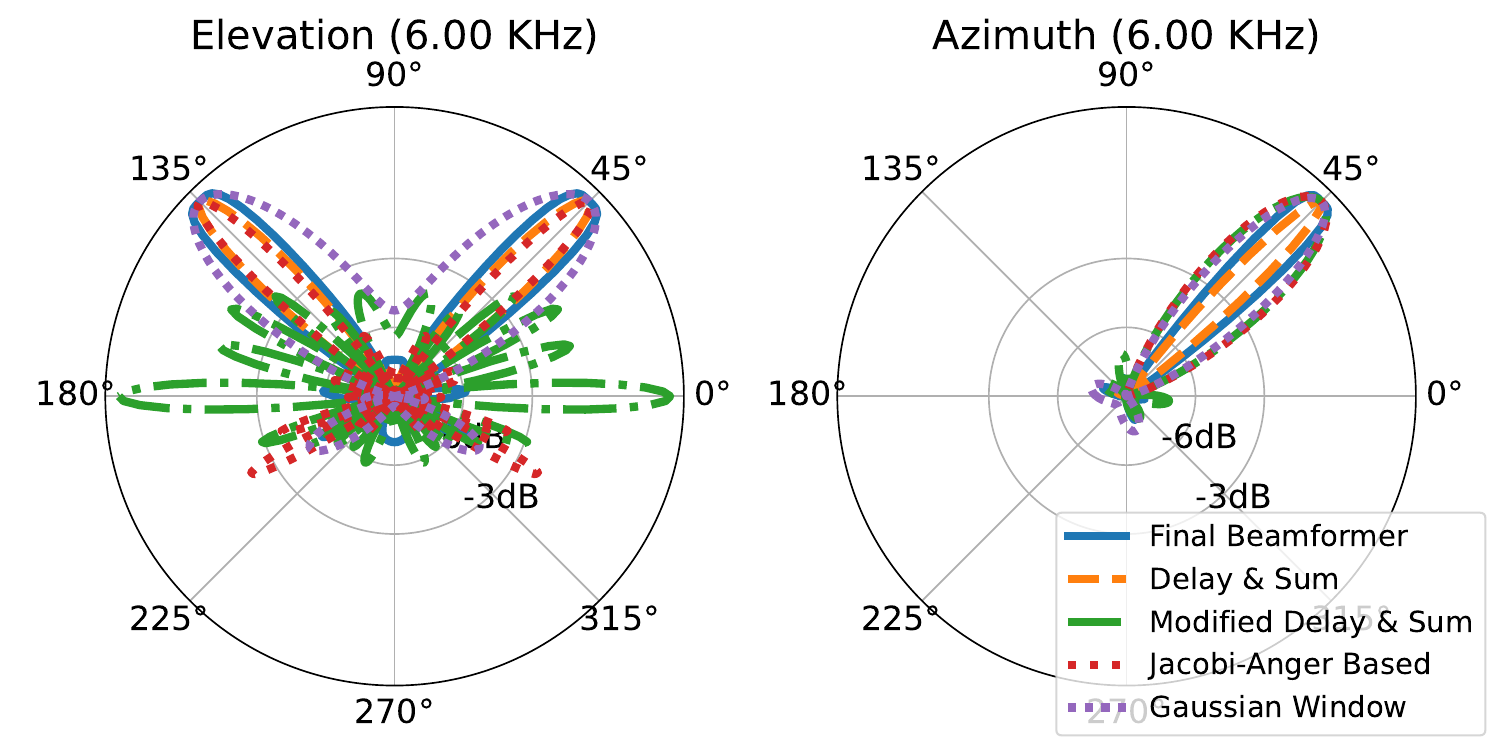}
     \vspace{-2mm}
     \caption{Comparison between the different methods tested.}
     \vspace{-2mm}
     \label{fig:comparison}
\end{figure}
We provide a comparative analysis of our proposed method against other widely used beamforming techniques. Specifically, we compare our method with the delay-and-sum beamformer, a modified delay-and-sum \cite{modified_ds}, a Jacobi-Anger expansion-based beamformer \cite{ccma_diff, insights}, and the modified gradient descent approach presented in \cite{isra3}, denoted as \com{Gaussian Window}. The resulting beampatterns for each method are illustrated in Fig.~\ref{fig:comparison}. This figure highlights the differences in spatial selectivity and mainlobe width achieved by each approach. The delay-and-sum beamformer serves as a conventional reference, providing a straightforward implementation but with limited control over the steering capabilities. As it can be seen in the figure, our method outperforms all the other approaches, specially in the elevation axis for lower frequencies.

\section{Conclusions and discussion}

In this work, we presented a framework that uses automatic differentiation to optimize the beamformer weights of a concentric circular microphone array (CCMA), allowing for controlled beamwidth in both elevation and azimuth. This approach enables the evaluation of different loss functions with various penalization terms, which simplifies the tuning process.

Our study also considers the inclusion of additional terms to promote performance invariance across the frequency range. The complete loss function comprises a performance term $P$, an invariance term $I$ that penalizes the standard deviation of metrics such as directivity factor (DF) and white noise gain (WNG) across frequencies, and a difference term $\Delta$ that addresses discrepancies between adjacent and opposing frequency bands. The parameters.

Our results indicate that the proposed method outperforms with conventional approaches and a modified gradient descent procedure. By considering both differential beamforming strategies and the benefits of the symmetric CCMA design, this work contributes to the ongoing effort to achieve frequency-invariant beamforming across a broad range of frequencies.

\section{Acknowledgments}
This work has been supported by Grants TED2021-131003B-C21 and TED2021-131401A-C22 funded by MCIN/AEI/10.13039/501100011033 and by the “EU Union NextGenerationEU/PRTR”, as well as by Grants PID2022-137048OB-C41 and  PID2022-137048OA-C43  funded by MICIU/AEI/10.13039/501100011033 and “ERDF A way of making
Europe”

\bibliography{ref}

@inproceedings{64MICMODULEOWN,
author = {Ortigoso Narro, Jorge and Moreno, Ricardo and de la Prida Caballero, Daniel and Raiola, Marco and Azpicueta-Ruiz, Luis},
year = {2024},
month = {09},
pages = {},
title = {64-MICROPHONE MODULE FOR A MASSIVE ACOUSTIC CAMERA}
}

@article{diff_theory1,
    author = {Chen, Jingdong and Benesty, Jacob and Pan, Chao},
    year = {2014},
    month = {Dec.},
    pages = {3097-3113},
    title = {On the design and implementation of linear differential microphone arrays},
    volume = {136},
    journal = {The Journal of the Acoustical Society of America},
    doi = {10.1121/1.4898429}
}

@book{diff_theory_book,
    author = {Benesty, Jacob and Chen, Jingdong and Pan, Chao},
    year = {2016},
    month = {05},
    pages = {},
    title = {Fundamentals of Differential Beamforming},
    isbn = {978-981-10-1045-3},
    doi = {10.1007/978-981-10-1046-0}
}

@INPROCEEDINGS{ccma_steer1,
  author={Lovatello, Jacopo and Bernardini, Alberto and Sarti, Augusto},
  booktitle={2018 26th European Signal Processing Conference (EUSIPCO)}, 
  title={Steerable Circular Differential Microphone Arrays}, 
  year={2018},
  volume={},
  number={},
  pages={11-15},
  doi={10.23919/EUSIPCO.2018.8553083}}

@article{Parra2006,
  title = {Steerable frequency-invariant beamforming for arbitrary arrays},
  volume = {119},
  ISSN = {1520-8524},
  url = {http://dx.doi.org/10.1121/1.2197606},
  DOI = {10.1121/1.2197606},
  number = {6},
  journal = {The Journal of the Acoustical Society of America},
  publisher = {Acoustical Society of America (ASA)},
  author = {Parra,  Lucas C.},
  year = {2006},
  month = jun,
  pages = {3839–3847}
}

@INPROCEEDINGS{ccma_diff,
  author={Huang, Gongping and Chen, Jingdong and Benesty, Jacob},
  booktitle={2018 IEEE International Conference on Acoustics, Speech and Signal Processing (ICASSP)}, 
  title={On the Design of Robust Steerable Frequency-Invariant Beampatterns with Concentric Circular Microphone Arrays}, 
  year={2018},
  volume={},
  number={},
  pages={506-510},
  doi={10.1109/ICASSP.2018.8461297}}

@INPROCEEDINGS{isragaussian,
  author={Sharma, Rajib and Cohen, Israel and Berdugo, Baruch},
  booktitle={ICASSP 2021 - 2021 IEEE International Conference on Acoustics, Speech and Signal Processing (ICASSP)}, 
  title={Window Beamformer for Sparse Concentric Circular Array}, 
  year={2021},
  volume={},
  number={},
  pages={4500-4504},
  doi={10.1109/ICASSP39728.2021.9414069}}

@ARTICLE{isra3,
  author={Peretz, Orel and Cohen, Israel},
  journal={IEEE/ACM Transactions on Audio, Speech, and Language Processing}, 
  title={Constant Elevation-Beamwidth Beamforming With Concentric Ring Arrays}, 
  year={2024},
  volume={32},
  number={},
  pages={1662-1672},
  doi={10.1109/TASLP.2024.3365390}}

@ARTICLE{israkroneken,
  author={Sharma, Rajib and Cohen, Israel and Berdugo, Baruch},
  journal={IEEE/ACM Transactions on Audio, Speech, and Language Processing}, 
  title={Controlling Elevation and Azimuth Beamwidths With Concentric Circular Microphone Arrays}, 
  year={2021},
  volume={29},
  number={},
  pages={1491-1502},
  doi={10.1109/TASLP.2021.3072275}}

@misc{autograd_beamformer_design,
  title     = {Applying Automatic Differentiation to Optimize Differential Microphone Array Designs},
  author    = {Siminfar, Samakoush Galougah and Duraiswami, Ramani},
  year      = {2024},
  eprint    = {2412.05123},
  archivePrefix = {arXiv},
  primaryClass  = {cs.SD},
  note      = {arXiv:2412.05123 [cs.SD]},
  url       = {https://arxiv.org/abs/2412.05123}
}

@article{tradeoff_kroneken_constant,
    author = {Frank, Ariel and Cohen, Israel},
    year = {2022},
    month = {05},
    pages = {829463},
    title = {Constant-Beamwidth Kronecker Product Beamforming With Nonuniform Planar Arrays},
    volume = {2},
    journal = {Frontiers in Signal Processing},
    doi = {10.3389/frsip.2022.829463}
}

@inproceedings{paszke2017automatic,
  title={Automatic differentiation in PyTorch},
  author={Paszke, Adam and Gross, Sam and Chintala, Soumith and Chanan, Gregory and Yang, Edward and DeVito, Zachary and Lin, Zeming and Desmaison, Alban and Antiga, Luca and Lerer, Adam},
  booktitle={NIPS-W},
  year={2017}
}

@INPROCEEDINGS{RProp,
  author={Riedmiller, M. and Braun, H.},
  booktitle={IEEE International Conference on Neural Networks}, 
  title={A direct adaptive method for faster backpropagation learning: the RPROP algorithm}, 
  year={1993},
  volume={},
  number={},
  pages={586-591 vol.1},
  doi={10.1109/ICNN.1993.298623}}

@ARTICLE{insights,
  author={Huang, Gongping and Chen, Jingdong and Benesty, Jacob},
  journal={IEEE/ACM Transactions on Audio, Speech, and Language Processing}, 
  title={Insights Into Frequency-Invariant Beamforming With Concentric Circular Microphone Arrays}, 
  year={2018},
  volume={26},
  number={12},
  pages={2305-2318},
  doi={10.1109/TASLP.2018.2862826}}

@INPROCEEDINGS{modified_ds,
  author={Yang, Y. and Chao Sun and Wan, C.},
  booktitle={Oceans 2003. Celebrating the Past ... Teaming Toward the Future (IEEE Cat. No.03CH37492)}, 
  title={Theoretical and experimental studies on broadband constant beamwidth beamforming for circular arrays}, 
  year={2003},
  volume={3},
  number={},
  pages={1647-1653 Vol.3},
  doi={10.1109/OCEANS.2003.178124}}

@article{source_localization,
    title = {Acoustic beamforming for noise source localization – Reviews, methodology and applications},
    journal = {Mechanical Systems and Signal Processing},
    volume = {120},
    pages = {422-448},
    year = {2019},
    issn = {0888-3270},
    doi = {https://doi.org/10.1016/j.ymssp.2018.09.019},
    url = {https://www.sciencedirect.com/science/article/pii/S088832701830637X},
    author = {Paolo Chiariotti and Milena Martarelli and Paolo Castellini}
}

@Article{windowbased,
    AUTHOR = {Long, Tao and Cohen, Israel and Berdugo, Baruch and Yang, Yan and Chen, Jingdong},
    TITLE = {Window-Based Constant Beamwidth Beamformer},
    JOURNAL = {Sensors},
    VOLUME = {19},
    YEAR = {2019},
    NUMBER = {9},
    ARTICLE-NUMBER = {2091},
    URL = {https://www.mdpi.com/1424-8220/19/9/2091},
    PubMedID = {31064067},
    ISSN = {1424-8220},
    DOI = {10.3390/s19092091}
}
\bibliographystyle{icml2025}


\end{document}